\documentclass[sigconf, nonacm]{acmart}

\usepackage{todonotes}

\renewcommand{\paragraph}[1]{%
  \vspace{0.5ex}%
  \noindent\textbf{#1.}%
}
\newcommand*\circled[1]{\textcircled{\raisebox{-0.9pt}{#1}}}

\begin{document}
\title{Selectivity Estimation for Semantic Filters on Image Data}

\author{Matthias Urban}
\affiliation{%
  \institution{Technical University of Darmstadt}
  \city{Darmstadt}
  \state{Hesse}
  \country{Germany}
}
\email{matthias.urban@tu-darmstadt.de}

\author{Vu Huy Nguyen}
\affiliation{%
  \institution{Technical University of Darmstadt}
  \city{Darmstadt}
  \state{Hesse}
  \country{Germany}
}
\email{vu_huy.nguyen@stud.tu-darmstadt.de}

\author{Gabriele Sanmartino}
\affiliation{%
  \institution{EURECOM}
  \city{Biot}
  \country{France}
}
\email{gabriele.sanmartino@eurecom.fr}

\author{Paolo Papotti}
\affiliation{%
  \institution{EURECOM}
  \city{Biot}
  \country{France}
}
\email{papotti@eurecom.fr}

\author{Carsten Binnig}
\affiliation{%
  \institution{TU Darmstadt, DFKI, HessianAI}
  \city{Darmstadt}
  \state{Hesse}
  \country{Gemany}
}
\email{carsten.binnig@tu-darmstadt.de}

\begin{abstract}
Semantic data systems integrate Large Language Models (LLMs) and Vision-Language Models (VLMs) directly into database query execution, enabling expressive queries on multi-modal data.
However, optimizing these queries requires accurate selectivity estimates to determine the most efficient operator execution order.
Contemporary systems rely on online sample-based profiling, a process that incurs severe latency overheads and struggles with low-selectivity queries.
In this paper, we introduce \emph{Semantic Histograms}, a novel selectivity estimator for semantic filters on image data that leverages shared embedding spaces to bypass traditional profiling.
We realize that all semantic filters are implicit range queries, as they match a range of different images.
Some filter predicates are more general, yielding a wide range, while others are more specific, yielding a smaller range.
To address the challenge of implicit ranges, we propose two approaches to estimate the queries' specificity, with an ensemble of the two performing best.
The evaluation shows that Semantic Histograms can reduce the end-to-end runtime overhead of query optimization and execution by up to 86\%.
\end{abstract}

\maketitle

\section{Introduction}

\paragraph{Query Optimization in Semantic Data Systems}
Recently, semantic data systems have emerged in both academia \cite{lotus, palimpzest, caesura, stretto, thalamusdb, abacus, docetl, docdb, flockmtl, continuous_prompts, samsara, aop, evergreen, zendb, galois} and industry \cite{aryn, cortex, bigquery}.
Fundamentally, these systems directly integrate Large Language Models (LLMs) and Vision Language Models (VLMs) into database query execution, enabling them to query large collections of multi-modal data.
For instance, in an e-commerce dataset, an analyst might search for product listings where a customer-uploaded image shows both a \emph{broken screen} and a \emph{warranty void sticker}. This query requires two distinct semantic filters to be evaluated by a VLM.

When optimizing such a query, the execution order of these filters significantly impacts runtime. If 90\% of the images feature a warranty void sticker, but only 0.1\% of images show broken screens, applying the \emph{broken screen} filter first is optimal as it minimizes the number of expensive model calls. The optimal execution depends on two main factors: (1) the latency of the VLM and (2) the selectivity of each semantic filter. While latency can be profiled offline, selectivity is highly query-dependent and must be determined at runtime.

\begin{figure*}
  \centering
  \vspace{-4ex}
  \includegraphics[width=0.85\linewidth]{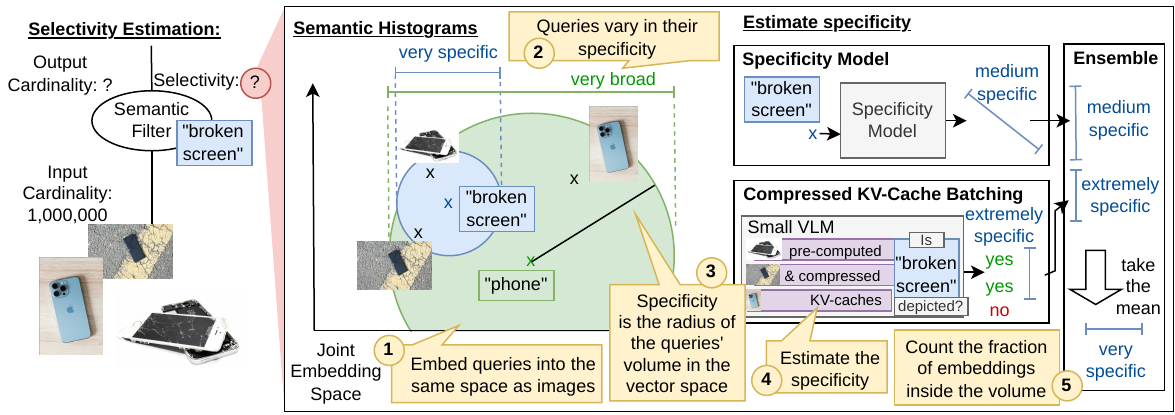}
  \caption{(Left) The semantic filter selects those pictures that depict a \emph{broken screen}. Thus, the filter selectivity is the fraction of images in the dataset that depict a broken screen. (Right) Semantic Histograms for selectivity estimation: \circled{1} we embed all images (offline) and the filter predicate \emph{broken screen} into a joint embedding space. High proximity in the vector space means a match is likely. \circled{2} Some predicates are more specific, intuitively covering a smaller volume in the vector space, while others are broader, covering a larger volume, like e.g., \emph{phone}. 
  To estimate the specificity (= the volume's radius \circled{3}) of a filter predicate, we propose three methods: \circled{4} The specificity model takes a predicate embedding as input and outputs a cosine distance threshold that is equivalent to the radius of the volume; if the distance between an image and a predicate embedding is smaller than it, they count as a match. The compressed KV-Cache batching approach optimizes LLM inference on a large set of images to obtain the actual VLM response for these images, which we use to compute a threshold as well. The ensemble averages the two thresholds to improve robustness. \circled{5} Finally, we count the matches and compute the selectivity.
  }
  \label{fig:overview}
  \vspace{-2ex}
\end{figure*}

\paragraph{The Overhead of Online Profiling}
Contemporary semantic data systems usually rely on an online profiling phase to estimate selectivity 
\cite{abacus, stretto, docdb, aop}.
In this phase, a small sample of data points (e.g., 1\% of images) is processed by the model to extrapolate the filter's behavior.
However, even a 1\% sample induces substantial overhead on large datasets; profiling a single filter on 1 million images would require 10,000 VLM calls.
Furthermore, sampling struggles to provide accurate estimates when the data is skewed or the target selectivity is low, leading to suboptimal query optimization decisions and additional overhead.

\paragraph{Semantic Histograms via Embedding Spaces}
To avoid these runtime overheads, traditional RDBMSs utilize histograms to estimate operator selectivity. Histograms are compact, offline-computed structures that capture a column's distribution, allowing the system to estimate filter selectivity without scanning the table. In this paper, we explore how to build a \emph{Semantic Histogram} for image data: a data structure that provides accurate selectivity estimates for visual semantic filters with low latency.

Traditional one-dimensional partitioning is inapplicable to inherently high-dimensional visual data.
A naive approach would pre-compute filter outcomes offline to populate standard histograms.
However, in semantic data systems, filter predicates are rarely known in advance; thus, this approach fails once the user applies new, unseen predicates.
Therefore, we base our Semantic Histogram on embedding spaces provided by models like CLIP or SigLIP \cite{clip, siglip, siglip2}. These models map images and text into a shared, high-dimensional vector space where semantic similarity corresponds to vector proximity. By calculating the cosine distance between a text filter predicate embedding (e.g., \emph{broken screen}) and the distribution of image embeddings, we can identify which and how many images most closely match the filter predicate.

\paragraph{Filter Predicates as Semantic Ranges}
However, turning these distances into accurate selectivities is non-trivial.
If we could define a strict distance threshold such that only images within the threshold match the filter predicate, we could easily estimate selectivity.
Yet, an optimal threshold depends heavily on the specific filter predicate. A helpful way to frame this is to view all filter predicates as \textbf{range queries} within the embedding space.
A broad filter predicate, such as \emph{phone}, conceptually covers a larger semantic volume (a wider range), meaning images further from the embedding are still relevant, thus requiring a higher distance threshold.
Conversely, a highly specific filter predicate, such as \emph{IPhone X with broken screen}, maps much closer to its target images due to the contrastive training of embedding models. 

\paragraph{Calibrating Thresholds for Semantic Histograms}
To overcome the challenge of unknown specificity,  we propose two methods for estimating the threshold.
In the evaluation, we demonstrate that an ensemble of the two methods works most robustly across datasets.

The first leverages a lightweight \emph{specificity model} trained on hierarchical visual labels to generalize across diverse filter predicates.
This model can directly map from an input filter-predicate embedding to a threshold with a fraction of the cost of a VLM call, yielding low-latency selectivity estimates that are often of high quality in our experiments.

The second estimates the threshold empirically.
It computes the actual VLM response on a large predefined sample and heavily optimizes the inference using a single, massive forward pass over pre-computed and compressed KV-caches of the images.
Using these intermediate image representations, the system can empirically ``peek'' at the data distribution at a fraction of the cost of standard inference. 
Moreover, instead of using the selectivity on the sample, we use it to compute a distance threshold, enabling more accurate predictions even when the sample contains no positive examples. We show that this approach has slightly higher latency than the first method, but yields more robust estimates.

\paragraph{Contributions and outline}
In summary, this paper makes the following contributions:
(1) \emph{Semantic Histograms:} We introduce a novel selectivity estimator for semantic filters on image data that leverages shared embedding spaces. To accurately capture filter-predicate specificity, we propose several threshold calibration techniques: a low-latency specificity model, a highly efficient compressed KV-cache batching method, and a robust ensemble of the two. All bypass the overhead of traditional online profiling. \\
(2) \emph{Empirical Evaluation:} We evaluate our proposed approaches across three image datasets from Caesura~\cite{caesura} and Sembench~\cite{sembench}, demonstrating that Semantic Histograms yield highly accurate selectivity estimates while significantly reducing query optimization latency compared to standard sample-based profiling.

The remainder of this paper is organized as follows: Section \ref{sec:overview} provides an overview of Semantic Histograms, and Section \ref{sec:specificity} details how we deal with predicates of different specificity.
Section \ref{sec:eval} presents our empirical evaluation, while related and future work, as well as the conclusion, are presented in Sections \ref{sec:related_work} and \ref{sec:conclusion}.

\section{Overview} \label{sec:overview}
In this section, we provide an overview of how we accurately estimate the selectivities of semantic filters on image data with low latency.
These selectivity estimates can be used to optimize semantic queries involving multiple semantic operators, as we demonstrate in Section \ref{sec:eval}.
The selectivity estimation is done in two phases: (1) offline, we pre-compute a synopsis of the available image data using embedding models such as CLIP~\cite{clip} or SigLIP~\cite{siglip, siglip2}, and (2) online, we use these embeddings for selectivity estimation.

\subsection{Offline Embedding Computation}
Offline, we need to capture the semantics of the images in our dataset in a way that allows for arbitrary filter predicates.
For instance, the user could query for \emph{phone with broken screen}, and we need to estimate the selectivity even when we have never seen that or a similar filter predicate before.
To do so, we use SigLIP2~\cite{siglip2} to first map all images into a high-dimensional embedding space that captures their semantics.
These enable us to predict the selectivity of arbitrary filter predicates, because filter predicates can be mapped into the same embedding space, where semantically similar images are close to the filter-predicate embedding.

In our Semantic Histogram, there is no concept of buckets as in regular histograms; we simply store all pre-computed embeddings as-is.
However, we did experiment with bucketizing the embeddings by fitting density estimators, such as Gaussian Mixture Models, to the embedding vectors, or by clustering them.
However, we found that simply keeping all embeddings yielded the best results: Their storage footprint is minimal (4.5 kB/embedding) compared to the original images (which need to be stored anyway), and computing the cosine similarity between a large number of embeddings is orders of magnitude faster than LLM calls (which happen anyway during query execution).
We found that bucketizing can sometimes reduce the accuracy of selectivity estimates, leading to suboptimal query optimization choices and runtime overhead.

\subsection{Online Selectivity Estimation}
Figure \ref{fig:overview} shows an overview of how Semantic Histograms compute selectivity estimates of semantic filters.
Selectivity is the fraction of images in the dataset that match the filter's predicate, e.g., the fraction of images that depict a \emph{broken screen} (Figure \ref{fig:overview}, left).
To estimate the selectivity, we extract the filter's predicate (e.g., \emph{broken screen}), and embed it into the same embedding space as the images, using SigLIP2 \cite{siglip2} again (See \circled{1} in Figure \ref{fig:overview}).

However, to decide how many image embeddings match the predicate embedding, we have to consider that the filter predicates differ in their \emph{specificity} (See \circled{2}).
For instance, a filter predicate such as \emph{broken screen} is more specific than a broader one such as \emph{phone}.
Intuitively, we want to map the filter predicates to a volume in the vector space (rather than a single point), and have all image embeddings within the volume match the predicate.
The specificity then corresponds to the radius of that volume (See \circled{3}); a highly specific predicate corresponds to a smaller radius, and a less specific predicate corresponds to a larger radius.
Thus, to decide how many image embeddings are inside the volume, we compute the cosine distances from the predicate embedding, which can be done orders of magnitude faster than calling an LLM.
An image embedding is inside the volume if it is smaller than a distance threshold equivalent to the volume's radius.

To estimate the distance threshold (i.e., the radius) of a filter predicate, we present two approaches (See \circled{4} in Figure \ref{fig:overview}).
Moreover, as we show in our evaluation, an ensemble of the two approaches yields particularly robust selectivity estimates across all datasets we tested.

The first approach is a simple, small neural network trained to estimate the specificities of filter predicates.
Its input is a filter predicate text embedding, and it directly outputs the distance threshold.
We find that this approach yields very fast estimates that are very accurate as long as the data at hand is somewhat similar to the specificity model's training data.

The second approach, compressed KV-cache batching, is more robust.
The idea is to select a large, diverse sample of images and optimize the inference on this sample as much as possible, so we can quickly decide which images in the sample match the filter predicate.
However, sampling notoriously struggles with low-selectivity filters.
Thus, instead of simply using the selectivity on the sample, which would be 0 if no image in the sample matches, we instead use it to compute a threshold.
That way, when there are no sample matches, we know it is a very specific filter predicate, resulting in a small distance threshold.
If any images match the filter predicate, they will likely be inside the volume defined by the threshold, leading to a strictly positive, and often more accurate, selectivity estimate.
We explain the details of the specificity model and the compressed KV-cache batching in Section \ref{sec:specificity}.

Finally, after estimating the distance threshold, we can use it to compute the filter selectivity (See \circled{5} in Figure \ref{fig:overview}).
We simply return the fraction of image embeddings that are closer to the filter predicate embedding than the threshold.

\section{The Specificity Challenge}\label{sec:specificity}
Estimating selectivities of semantic filters comes with a unique challenge:
Some filter predicates are more specific, like for instance \emph{IPhone X with broken screen}, and others are less specific, like for instance \emph{phone}.
In this section, we explain how to estimate the specificity of a filter predicate, which is essential for computing filter selectivities, as explained before.
All approaches output a cosine distance threshold as a measure of specificity.
A small distance threshold leads to the predicate covering a smaller volume in the vector space, and thus corresponds to more specific filter predicates.
Conversely, a large distance threshold corresponds to less specific predicates.
The presented approaches differ in their strengths.
However, our final ensemble approach, in particular, yields very accurate and robust selectivity estimates, as shown in our evaluation.

\subsection{Specificity Model}

The specificity model is a small neural network that maps a predicate embedding to its distance threshold.

\paragraph{Model Training} However, training the model is challenging.
For training it, we need a large dataset that pairs text embeddings with distance threshold labels.
To this end, we use ImageNet \cite{imagenet}, which is a large collection of labeled images.
Importantly, each image has not only a single label but a hierarchy of labels, inherited from Wordnet \cite{wordnet}, ranging from specific concepts (e.g., cellular telephone) to broader concepts (e.g., telephone, electronic device).
We use a subset consisting of 150,000 photographs divided into 1000 categories \cite{imagenet-subset}.
This allows us to construct a specificity dataset as follows:
First, we randomly sample a subset of the data and several WordNet concepts that appear in it.
For instance, we might have sampled the concepts of \emph{cellular phone} and \emph{electronic device}.
Then we compute the label threshold for each concept.
For instance, for \emph{cellular phone}, we have 10 images in our subset that match that label.
Then we compute the threshold label for training, such that exactly 10 image embeddings in the subset are closer to the embedding of \emph{cellular phone} than the threshold.
We repeat this process until we have more than 5000 training samples.

\paragraph{Limitations} While this leads to very fast threshold estimates, it has limitations stemming from the dataset choice.
ImageNet focuses heavily on animals (especially different dog breeds), and the concepts are often scientific.
In the future, we want to fine-tune the embedding models themselves to return not only embeddings but also the corresponding similarity thresholds, using a much broader dataset spanning more domains than ImageNet.

\begin{figure}
  \centering
  \vspace{-4ex}
  \includegraphics[width=0.8\linewidth]{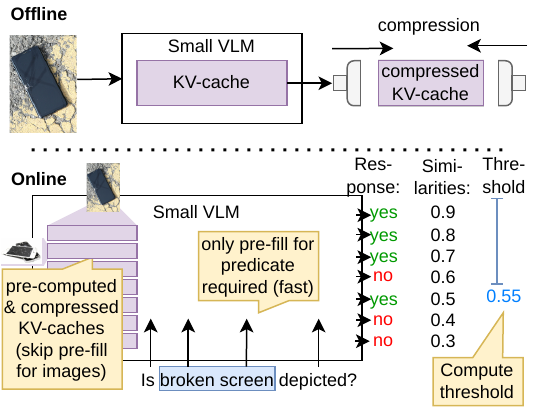}
  \caption{(Top) Offline, we feed a selected number of images into a small VLM (e.g., LLaVA-NeXT 8B \cite{llava-next}) and extract and compress the KV-cache. Afterward, we load the compressed KV-caches onto the GPU. (Bottom) Online, a filter predicate comes in. We can skip the prefill phase for the images and quickly compute the responses for all selected images at once. We also compute the similarities between the selected images and the predicate to compute a similarity threshold, which we then use for selectivity estimation.}
  \label{fig:kv-batching}
  \vspace{-2ex}
\end{figure}

\subsection{Compressed KV-cache Batching}
While the first method yields fast estimates that are accurate when the dataset at hand does not deviate too much from ImageNet, we also offer a more robust method.
To do so, we compute the actual VLM responses on a well-chosen subset of images to determine the threshold and make the VLM inference as fast as possible.

\begin{figure*}
  \centering
  \vspace{-4ex}
  \includegraphics[width=\linewidth]{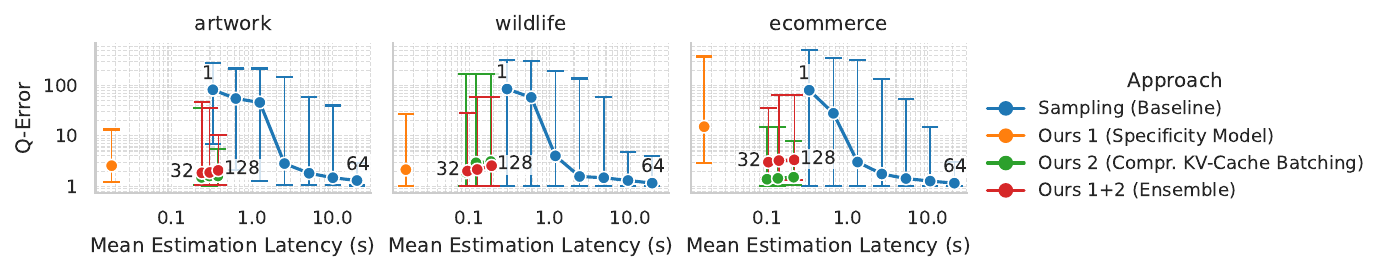}
  \caption{Semantic Histograms are much faster than sampling, while being very accurate. We plot the Q-Error (Y-axis) of different selectivity estimation approaches and configurations against their latency (X-axis). Good approaches are at the bottom (low Q-Error) and on the left (low latency) of each plot. For latency, we plot the mean across all filter predicates and seeds; for Q-Error, we plot the median and the 5th and 95th percentiles (as error bars). For sampling, we use sample sizes of 1, 2, 4, \dots, 64; for compressed KV-Cache batching, we use sample sizes of 32, 64, 128 (with compression rates of 0.6, 0.8, 0.9, respectively, to keep memory consumption approximately the same). We show which configuration is used next to the markers.
  }
  \label{fig:qerror}
  \vspace{-2ex}
\end{figure*}

\paragraph{Threshold estimation} 
To estimate the threshold, we ask the VLM whether, e.g., a broken screen is depicted on a sample of images.
We also compute the similarities of \emph{broken screen} with all images in the sample.
If, for instance, 10 images depict a broken screen, we choose the distance threshold so that exactly 10 images in the sample have a cosine distance smaller than the threshold.
If no image in the sample matches the filter predicate, we set the threshold to the smallest observed distance, allowing us to still estimate selectivity for very low-selectivity predicates.

\paragraph{Making inference fast}
However, if done naively, this does not lead to faster estimates than the online profiling phase typically used, which we want to eliminate.
Thus, we optimize the inference to get many responses (in our case, up to 128) in a fraction of the time that typical inference would normally take for a single response.
To do so, we first fix the sample to 128 images as explained in the next paragraph.
Then, inspired by Stretto \cite{stretto}, we feed these images offline into the LLM and precompute and compress the model's KV cache, as shown in Figure \ref{fig:kv-batching} (top).
Online, this allows us to skip image processing with the VLM by omitting the so-called prefill phase, see Figure \ref{fig:kv-batching} (bottom).
Moreover, we compress these KV-caches using the Expected Attention press \cite{expected_attention}, so that all 128 embeddings take about 4 GB of space with 90\% compression in our experiments.
Thus, they can be preloaded onto the GPU before any query enters the system.
Compression is lossy and degrades prediction quality, but allows for a large sample to be pre-loaded onto the GPU, leading to lower estimation latencies, which is beneficial.

When the query, e.g., with the filter \emph{broken screen}, arrives, we only need two more VLM passes to obtain responses for all 128 images in a batch.
First, we need to finish the prefill phase by computing the KV-cache of the prompt (i.e., Is \emph{broken screen} depicted?), which usually consists of only a few tokens and is thus very fast.
Afterward, we generate a single yes/no answer token for all images at once.
Experiments show that this method yields robust threshold estimates with latencies smaller than a single LLM call.

\paragraph{Sample Selection}
Finally, it is important that the sample used to estimate the threshold is diverse and representative, capturing many of the concepts present in the dataset.
To select a diverse sample, we cluster the image embeddings using K-Means.
We set the number of clusters to the sample size we want to obtain, e.g., up to 128 in our case.
Then, we pick the images whose embeddings are closest to each centroid.
Finally, we want to mention that this method also allows for larger sample sizes.
However, it would consume more memory (about 30~MB per KV-cache using LLaVA-NeXT 8B~\cite{llava-next} with 90\% compression), potentially leaving some KV-caches not pre-loaded onto the GPU, requiring them to be loaded online.

\subsection{Ensemble}
Finally, we found that the two threshold estimation approaches make orthogonal mistakes.
Combining the two approaches by averaging the two predicted thresholds often yields better selectivity estimates than either approach alone, as shown in our experiments.

\section{Evaluation}\label{sec:eval}
In our experiments, we show that the Semantic Histogram variants presented in this paper are both faster and more accurate than estimating selectivity via sampling in an online profiling phase, as is done by contemporary semantic data systems.
Especially the ensemble yields robust estimates across all datasets with low latency, resulting in significantly faster end-to-end runtimes.

\subsection{Experimental Setup}

\paragraph{Datasets} As our datasets, we use three datasets introduced by prior work \cite{sembench, caesura} to evaluate semantic data systems.
All contain a column of image data, which we can use to apply semantic filters to images and, more importantly, estimate the selectivities of these filters.
\emph{(1) Artwork:} A dataset constructed from Wikidata, that contains 1000 pictures of artworks from different epochs. The dataset was first introduced by Caesura \cite{caesura}.
\emph{(2) Wildlife:} A dataset containing wildlife photographs of different animals from a conservancy in Kenya. We randomly select a subset of 1000 images. The dataset was first introduced by SemBench \cite{sembench}.
\emph{(3) E-Commerce:} A dataset containing pictures of different fashion items. We select a subset of 1000 images. First introduced by Sembench \cite{sembench}.

\paragraph{Queries and Operators}
For each dataset, we generate a large set of filter predicates (14-26 per dataset) based on the original queries. We also ensure that we have predicates of different specificity; for instance, in the wildlife dataset, we have filter predicates such as \emph{animal}, \emph{prey animal}, \emph{antelope}, and \emph{impala}.

For executing the semantic filters, we use the Qwen 2.5 VL 7B model~\cite{qwen-2-vl, qwen-2.5-vl}, and the prompt \emph{Is <filter predicate> depicted?} We deploy the model using ollama on an A100 GPU.

\paragraph{Metrics}
To evaluate the quality of selectivity estimates, we use the Q-Error, which is the ratio of the predicted selectivity to the actual selectivity.
For instance, if the true selectivity is 2\% but a method predicts 20\% or 0.2\%, the Q-Error is 10 in both cases.
To avoid undefined Q-errors when the predicted selectivity is 0, we set the predictions to $\frac{1}{\mathrm{dataset \_size}}$ in this case.
Moreover, we measure the latency of the different selectivity estimation methods and the time to execute semantic queries in seconds.

\paragraph{Approaches and Baselines}
We test the different Semantic Histogram approaches presented and compare them with sampling in an online profiling phase for selectivity estimation.
For sampling, we process a data sample with the VLM and estimate selectivity based on its responses.
Note that we picked a small 7B VLM; a larger one would exacerbate sampling costs.
We test a sample size of 1 (0.1\% of dataset size), 2, 4, \dots, 64 (6.4\% of dataset size).

Moreover, for the compressed KV-cache batching approach, we test several configurations that use the same amount of GPU memory: (1) we pre-compute the KV-cache of 128 images and compress them with a compression ratio of 90\% (2) we pre-compute 64 images and use a compression ratio of 80\%, and (3) we pre-compute 32 KV-caches and use a compression ratio of 60\%.
We use LLaVA-NeXT 8B~\cite{llava-next}, where the KV-caches for each setting are about 4 GB. 
We run all experiments with 20 seeds and average the results.

\begin{figure*}
  \centering
  \includegraphics[width=\linewidth]{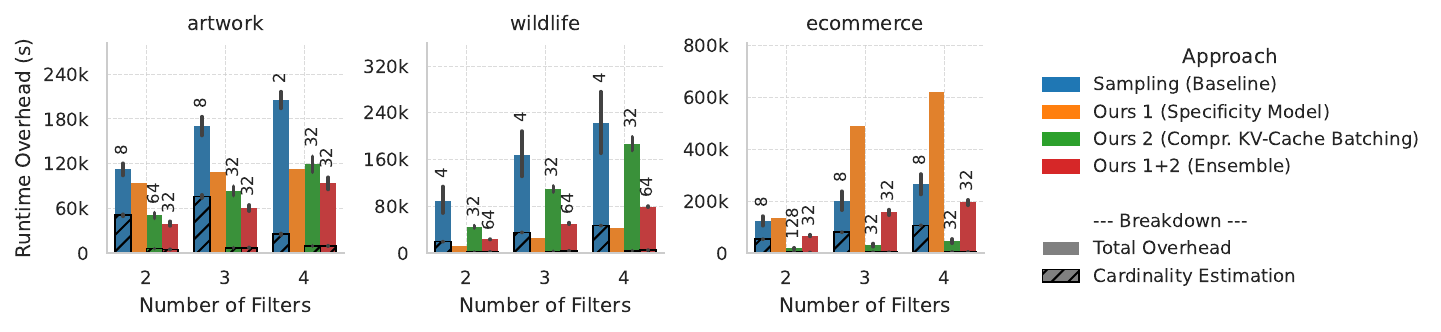}
  \vspace{-4ex}
  \caption{Using Semantic Histograms for query optimization leads to faster end-to-end runtime than sampling.
  We generate semantic queries with 2, 3, and 4 semantic filters (Y-axis) and plot the absolute runtime overhead of query optimization + execution when using the selectivity estimation methods.
  For each approach, we plot the mean overhead and its 95\% confidence interval (error bars) for the best-performing configuration (e.g., the best-performing sample size, annotated above each bar) across different seeds.
  The overhead is the difference from a perfect baseline that uses a zero-latency oracle selectivity estimation.
  The ensemble approach for Semantic Histograms is most robust across datasets and works better than sampling on all datasets.
  }
  \label{fig:reorder}
  \vspace{-2ex}
\end{figure*}

\subsection{Fast and Accurate Selectivity Estimates}
\paragraph{Summary}
In our first experiment, we show that Semantic Histograms yield better selectivity estimates in a fraction of the estimation latency.
The ensemble, in particular, yields excellent Q-errors across all datasets, with latencies lower than a single LLM call.
The other Semantic Histogram variants are less robust but excel on particular datasets; for instance, the specificity model performs well on the wildlife dataset, likely because it contains pictures of animals, which also appear frequently in ImageNet.
Compressed KV-Cache batching works excellently on e-commerce and is also quite robust.

\paragraph{Details}
In Figure \ref{fig:qerror}, we plot the Q-Error of selectivity estimation (Y-axis)  against the latency to obtain the estimates (X-axis) of the different approaches.
Low latency is important as any latency for estimating selectivity directly translates into runtime overhead.
Low Q-Error is important for enabling good decisions when selectivity estimates are used, for instance, in query optimization.
Even minor mispredictions can cause the query optimizer to pick the wrong operator order, resulting in runtime overhead.

First, we observe that all our approaches are substantially faster than naive \emph{sampling} during an online profiling phase, depicted in blue.
In particular, the \emph{specificity model}, which does not involve any LLM call for selectivity estimation, is significantly faster than all other approaches.
On average, it needs only 17 ms to make a prediction.
But also, the \emph{compressed KV-cache batching} approach and the \emph{ensemble} are faster than sampling.
For instance, when pre-loading 128 KV-caches onto the GPU and computing the LLM response for 128 samples in a single batch, we get these 128 responses in approximately the same time as it takes to evaluate a single sample.

At the same time, our predictions are often more accurate than sampling with reasonable sample sizes.
In particular, we want to highlight the ensemble, which achieves good median Q-error scores of 1.8, 2.0, and 3.0 on the datasets and is particularly robust.
Sampling needs to process about 1\% of data with the VLM to obtain the same median (and p95 Q-error), which incurs substantial runtime overhead.
The other approaches excel on particular datasets.
For instance, the specificity model works very well on wildlife, likely because it is similar to ImageNet, which also contains many animals.
On the other hand, the compressed KV-cache batching excels on the e-commerce dataset.
We speculate that this is because these images are low-complexity, each showing a single product, enabling accurate predictions even with KV-cache compression.

As expected, the specificity model does not work well on all datasets.
In particular, on e-commerce, we observe a high median Q-Error of 15.2, which is due to threshold mispredictions.
However, ensembling helps to improve the Q-Error.
Another observation worth noting is that the compressed KV-Cache Batching often predicts very low selectivity for wildlife, even for rather broad concepts such as \emph{prey animal}, leading to a high p95 Q-Error.
Upon investigation, we found that this is a problem with the underlying VLM we use (LLaVA-NeXT 8B), which is not very accurate on these images (potentially because the animals are often very small and in the distance), rather than a side-effect of KV-cache compression loss.
Thus, we suspect that implementing this technique with more accurate models will yield more accurate selectivity estimates.
However, again, ensembling helps reduce the p95 Q-Error.

\subsection{End-to-End Runtime}
\paragraph{Summary}
In our second experiment, we show that the improved selectivity estimates with low latency lead to better end-to-end runtimes when used for query optimization.
The ensemble yields the lowest overall end-to-end runtimes, with statistically significant speedups over naive sampling on all datasets.

\paragraph{Details}
To understand the trade-off between accuracy and latency in selectivity estimation for semantic data systems, we use the selectivity estimates for query optimization and measure end-to-end runtime.
In this setting, there are two types of overhead a bad selectivity estimator can have: (1) the latency for producing the selectivity estimate (2) the latency induced by producing bad estimates leading to a suboptimal plan.
Thus, to achieve low end-to-end latency, the selectivity estimator must be both fast and accurate.

The experiment setup is as follows:
For each dataset, we generate 300 semantic queries - 100 each of 2, 3, or 4 consecutive semantic filters, randomly sampled from the available filter predicates.
During query optimization, we must determine the filter order.
Ideally, we want to have the most selective filter first, so that we minimize the number of images the second filter must process, and so on.
Thus, for each query and selectivity estimator, we estimate the selectivity of each filter in a query.
Then we sort the filters by this value, putting the most selective filter first.
Afterward, we run all filters on the full dataset in this order.

In Figure \ref{fig:reorder}, we plot the sum of overheads over all queries of the end-to-end runtime versus an optimal query optimizer that has access to oracle selectivity estimates with zero latency.
Overall, we see that our Semantic Histogram approaches are typically faster than sampling, with only a few exceptions.
Moreover, the ensemble method is better than sampling across all datasets and filter counts.
Usually, the error bars, which denote the confidence intervals of the mean overhead across different seeds, do not overlap with those of sampling, indicating that the result is statistically significant.
Again, the other Semantic Histogram variants excel on individual datasets: the specificity model excels on wildlife, and the compressed KV-cache batching on e-commerce.
Finally, the error bars of sampling are much larger than those of Semantic Histograms, indicating that the runtime of systems relying on sampling is highly unpredictable.

\section{Related Work}\label{sec:related_work}
\paragraph{Selectivity Estimation}
Traditionally, histograms or sketches have been used for selectivity estimation \cite[e.g.,][]{system-r, histograms, hyperloglog}.
More recently, also machine learning-based approaches have been explored \cite[e.g.,][]{deepdb, mscn, naru}.
However, these approaches require the fixed schema provided by traditional RDBMSs.
Concurrent work by \citet{ce-semantic-text} has explored how to build selectivity estimators for text data in semantic data systems.
It does not address image data.

\paragraph{Semantic Data Systems}
Recently, many semantic data systems have been proposed \cite{lotus, palimpzest, caesura, stretto, thalamusdb, abacus, docetl, docdb, flockmtl, continuous_prompts, samsara, aop, evergreen, aryn, cortex, bigquery}. Many include query optimizers that pick the best operator order \cite{abacus, palimpzest, stretto, docdb, aop, bigquery, cortex}, pick the model or combination of models with the best quality-runtime trade-off \cite{lotus, palimpzest, abacus, stretto, cortex, bigquery}, or pick the best prompting strategies \cite{abacus}, or rewrite the queries for better quality or lower latency \cite{docetl, moar, cortex}.
However, these approaches rely on sampling when they estimate selectivities of semantic filters.

\section{Conclusion and Future Work}\label{sec:conclusion}
In this paper, we show that image embeddings can be used to build selectivity estimators for image data that yield low-latency and high-quality estimates.
These can be used by semantic data systems for query optimization to achieve a lower end-to-end runtime than previous approaches that use sampling for selectivity estimation.
In the future, we want to extend the approach towards other modalities, such as text or audio, and also other semantic operators, such as semantic joins.
Moreover, some semantic data systems rely on sampling to estimate the \emph{quality} of different operator implementations (e.g., using different LLM sizes).
Therefore, future work must also look into how sampling can be replaced for these purposes.


\begin{acks}
This work was funded by the DFG/ANR Project MAgiQ (ANR-24-CE92-0077; DFG, German Research Foundation – Project No. 545611510), the LOEWE Spitzenprofessur programme (III 5-519/05.00.003-(0005)), by the Deutsche Forschungsgemeinschaft (DFG, German Research Foundation) under Germany’s Excellence Strategy (EXC-3057/1 “Reasonable Artificial Intelligence”, Project No. 533677015), and by the French government, through the 3IA Côte d’Azur Investments in the IA-cluster project managed by the National Research Agency (ANR-23-IACL-0001). We also thank DFKI and hessian.AI.
\end{acks}

\balance
\bibliographystyle{ACM-Reference-Format}
\bibliography{sample}

\end{document}